\documentclass[journal,onecolumn]{IEEEtran}
\ifCLASSINFOpdf
\else
\fi

	\usepackage{amssymb}
	\usepackage{graphicx}%
	\usepackage{amsmath,amssymb,amsfonts}%
	\usepackage{amsthm}%
	\usepackage{mathrsfs}%
	\usepackage{xcolor}%
	\usepackage{physics}
	\usepackage{hyperref}
	\newcommand{\bLozenge}{\mathbin{\blacklozenge}}
	
	\theoremstyle{thmstyleone}%

	\theoremstyle{thmstyletwo}%
	\newtheorem{example}{Example}%

	\theoremstyle{thmstylethree}%
\begin{document}
		\title{Quantum Key Distribution Based on Systematic Polar Coding}
		
		\author{Georgi~Bebrov
			
			% <-this % stops a space
			\thanks{G. Bebrov was with the Department
				of Telecommunications, Technical University of Varna, Varna, 9010 BULGARIA e-mail: g.bebrov@tu-varna.bg.}% <-this % stops a space
			% <-this % stops a space
			}
		

	\markboth{Journal of \LaTeX\ Class Files}%
	{Shell \MakeLowercase{\textit{et al.}}: Bare Demo of IEEEtran.cls for IEEE Journals}



	\maketitle
	
	\begin{abstract}
		Here we concerned with quantum key distribution\textemdash a way to establish common cryptographic key between several parties. The work proposes a combination between quantum key distribution and systematic polar coding (an error correction algorithm) frameworks\textemdash quantum key distribution based on systematic polar coding. This results in obtaining key rates greater than standard quantum key distribution (BB84) and its efficient version (eBB84) when finite-size regime and lower-error-rate quantum channel are considered.
	\end{abstract}
	

	% \fi
	\IEEEpeerreviewmaketitle
	
	\section{Introduction}\label{sec1}
	\IEEEPARstart{I}{n} the information age, the necessity of a secure communication and storage of sensitive information is of utmost importance\textemdash bank transfers, healthcare data, military data, and so on. A way to provide  an information-theoretic security of a communication process is by using primitives (algorithms, methods, protocols) of the so-called \textit{quantum cryptography}. The most prominent representative of this scientific/research field is the \textit{quantum key distribution} (QKD) \cite{Bennett1984,Ekert1991,Lo2005,Lo2012,Lucamarini2018,Wang2018,Curty2019}. This primitive enables two or more parties to establish a common cryptographic key by means of \textit{quantum phenomena}, such as \textit{uncertainty principle} \cite{Bennett1984} or/and \textit{quantum entanglement} \cite{Ekert1991}. The most notable primitive is the so-called \textit{BB84-QKD} \cite{Bennett1984}, the original quantum key distribution model (protocol). \\
	\indent A long-standing open problem in the domain of quantum key distribution is developing protocols (schemes, model) with as high as possible key rates. As shown in \cite{Lo2005}, a way to obtain a higher-rate QKD is to use bias choice for both preparation and measurement bases. Such a QKD model is known as \textit{efficient} BB84-QKD (eBB84-QKD). This approach demonstrates its advantage to the full extent only in the case of an asymptotic regime of operation ($N\rightarrow\infty$, where $N$ signifies the amount of qubits in a single run of a QKD). In the case of a finite-size regime (finite $N$), the bias approach demonstrates results being close to those of a standard QKD (\textit{e.g.}, BB84-QKD). \\
	\indent In this regard, this work proposes a \textit{protocol process} (a series of protocol runs) that exhibits a key-rate behavior better than BB84-QKD and eBB84-QKD when finite-size regime and lower-error-rate quantum channel are considered. The proposal is based on the use of systematic polar coding \cite{Arikan2011}\textemdash a channel coding (error correction) method. The polar coding is incorporated into the transmission (quantum) stage of a QKD protocol (more precisely, BB84-QKD) in order for the error correction (information reconciliation \cite{Brassard1994}) procedure to be avoided. This way the key-rate reduction due to information reconciliation of a quantum key distribution is mitigated\textemdash higher key rates are attainable. \\
	\indent The paper is organized, as follows. Section \ref{polar-coding} recalls in a brief manner the concept of polar coding. Section \ref{pc-qkd} introduces: (i) a quantum key distribution protocol process based on systematic polar coding, called for short \textit{Polar-code} QKD; (ii) a key-rate evaluation of the newly-introduced QKD; (iii) a key-rate comparison between Polar-code QKD, BB84-QKD, and eBB84-QKD. Section \ref{conclusion} puts forward the conclusion of this work.

	%\IEEEpubidadjcol
	
	\section{Polar coding}\label{polar-coding}
	\textit{Polar coding} (\textit{code}) is invented by E. Ar{\i}kan \cite{Arikan2009,Arikan2011}. It is a linear block channel coding scheme proved to have a capacity-reaching behavior. The polar coding is given by the following transformation \cite{Arikan2011}
	\begin{equation}\label{polar-transf}
		\mathbf{x} = \mathbf{uG}
	\end{equation} 
	where $\mathbf{u}$ is the source sequence (undergoing polar coding), $\mathbf{x}$ is the codeword (encoded) sequence, and $\mathbf{G}$ is the polar-coding operator. Note that $\mathbf{u}$,$\mathbf{x}$ $\in$ $\mathbb{F}^N$ and $\mathbf{G}$ $\in$ $\mathbb{F}^{N\cross N}$, where the field $\mathbb{F}$ is usually the binary field $\mathbb{F}_2$. The variable $N$ denotes the size of both source and encoded sequences ($N$ $\in$ $\{2^n; n\in\mathbb{Z}_{+}\}$, $N>>1$). The polar-coding operator has the form \cite{Arikan2011}
	\begin{equation}
		\mathbf{G} = \mathbf{F}^{\otimes n}
	\end{equation}
	where $\otimes$ denotes the so-called \textit{Kronecker} (\textit{tensor}) \textit{power} (\textit{product}) and $\mathbf{F}$ is characterized by the matrix representation
	\begin{equation}
		\mathbf{F} = \begin{bmatrix}
			1 & 0\\
			1 & 1
		\end{bmatrix}
	\end{equation}
	The source sequence $\mathbf{u}$ is comprised of two parts: \textit{data bits} (symbols) $\mathbf{d}$ $\in$ $\mathbb{F}^K$ and \textit{frozen bits} $\mathbf{f}$ $\in$ $\mathbb{F}^F$. Note that all the frozen bits are usually set to $0$ (any other fixed choice on the frozen bits is also valid). The variables $K$ and $F$ are related, as follows:\textemdash
	\begin{equation}
		K + F = N
	\end{equation}
	This relation can be represented in the form
	\begin{equation}
		[1-H(E)]N + [H(E)]N = N
	\end{equation}
	$E$ being the error rate of the channel over which the sequence $\mathbf{x}$ travels in order to reach a recipient. Given the condition $N$ $\in$ $\{2^n,n\in\mathbb{Z}_{+}\}$, we should emphasize on that 
	\begin{equation}
		F = \lceil H(E)N \rceil
	\end{equation}
	and
	\begin{equation}
		K = N - F
	\end{equation}
	We now consider a \textit{systematic polar code} \cite{Arikan2011}. By definition, it is a sequence of encoded bits which is comprised of encoded frozen bits and data bits being in their initial/plain form. A systematic polar code can be obtained the following way. For the sake of clarity, a short example of systematic polar coding is worked out. 
	\begin{example}
		Suppose $N = 8$ and $E = 0.11$. Then $F = \lceil H(E)N \rceil$ $=$ $\lceil 0.4999N \rceil$ $=$ $4$ and $K = N - F = 4$. The source sequence $\mathbf{u}$ has the form
		\begin{equation*}
			\mathbf{u} = \begin{bmatrix}
				u_1 & u_2 & u_3 & u_4 & u_5 & u_6 & u_7 & u_8
			\end{bmatrix}
		\end{equation*}
		Suppose further that the frozen bits positions are $1$, $2$, $3$, and $5$, and all frozen bits are set to $0$. That is,
		\begin{eqnarray*}
			\mathbf{u} &&= \begin{bmatrix}
				f_1 & f_2 & f_3 & d_1 & f_4 & d_2 & d_3 & d_4
			\end{bmatrix}\nonumber\\
			&&= \begin{bmatrix}
				0 & 0 & 0 & d_1 & 0 & d_2 & d_3 & d_4
			\end{bmatrix}
		\end{eqnarray*}
		Notice that $\mathbf{f} = \begin{bmatrix} f_1 & f_2 & f_3 & f_4\end{bmatrix}$ is the sequence of frozen bits and $\mathbf{d} = \begin{bmatrix} d_1 & d_2 & d_3 & d_4\end{bmatrix}$ is the sequence of data bits.\\
		\indent The polar-coding operator $\mathbf{G}$, for $N=8$, has the form
		\begin{eqnarray*}
			\mathbf{G} &&= \mathbf{F}^{\otimes n} = \mathbf{F}^{\otimes 3} = \begin{bmatrix}
				1 & 0\\
				1 & 1
			\end{bmatrix} \otimes
			\begin{bmatrix}
				1 & 0\\
				1 & 1
			\end{bmatrix} \otimes
			\begin{bmatrix}
				1 & 0\\
				1 & 1
			\end{bmatrix} \nonumber \\
			&&= \begin{bmatrix}
				1 & 0 & 0 & 0 & 0 & 0 & 0 & 0\\
				1 & 1 & 0 & 0 & 0 & 0 & 0 & 0\\
				1 & 0 & 1 & 0 & 0 & 0 & 0 & 0\\
				1 & 1 & 1 & 1 & 0 & 0 & 0 & 0\\
				1 & 0 & 0 & 0 & 1 & 0 & 0 & 0\\
				1 & 1 & 0 & 0 & 1 & 1 & 0 & 0\\
				1 & 0 & 1 & 0 & 1 & 0 & 1 & 0\\
				1 & 1 & 1 & 1 & 1 & 1 & 1 & 1\\
			\end{bmatrix}
		\end{eqnarray*}
		Hence the codeword (encoded) sequence $\mathbf{x}$ becomes, based on Equation \eqref{polar-transf},
		\begin{eqnarray*}
			\resizebox{.9\hsize}{!}{$
				\begin{bmatrix}
					(d_1\oplus d_2\oplus d_3\oplus d_4) & (d_1\oplus d_2\oplus d_4) & (d_1\oplus d_3\oplus d_4) & (d_1\oplus d_4) & (d_2\oplus d_3\oplus d_4) & (d_2\oplus d_4) & (d_3\oplus d_4) & (d_4)
				\end{bmatrix} $}
		\end{eqnarray*}
		Notice that all the additions and multiplications above are performed in $\mathbb{F}_2$. In order to yield a systematic codeword sequence, we replace the encoded data positions by their plain (original) form, as follows:\textemdash
		\begin{eqnarray*}
			&&\mathbf{x}=\resizebox{.9\hsize}{!}{$
				\begin{bmatrix}
					(d_1\oplus d_2\oplus d_3\oplus d_4) & (d_1\oplus d_2\oplus d_4) & (d_1\oplus d_3\oplus d_4) & (d_1\oplus d_4) & (d_2\oplus d_3\oplus d_4) & (d_2\oplus d_4) & (d_3\oplus d_4) & (d_4)
				\end{bmatrix} $} \\ 
			&&\text{systematization of $\mathbf{x}$}\; \Longrightarrow \\
			&&\mathbf{x}'=\resizebox{.9\hsize}{!}{$
				\begin{bmatrix}
					(d_1\oplus d_2\oplus d_3\oplus d_4) & (d_1\oplus d_2\oplus d_4) & (d_1\oplus d_3\oplus d_4) & (d_1) & (d_2\oplus d_3\oplus d_4) & (d_2) & (d_3) & (d_4)
				\end{bmatrix} $}
		\end{eqnarray*}
		where $\mathbf{x}'$ is the systematic codeword sequence. Note that, in a systematic polar coding, only the frozen bits undergo an actual encoding process.
	\end{example}
	
	\section{Polar-code QKD}\label{pc-qkd}
	In this section, a single-photon point-to-point quantum key distribution protocol based on systematic polar coding (Polar-code QKD) is proposed. Moreover its performance is examined.
	\subsection{Protocol description}
\begin{figure}[h]
		\centering
		
		\includegraphics[scale=1]{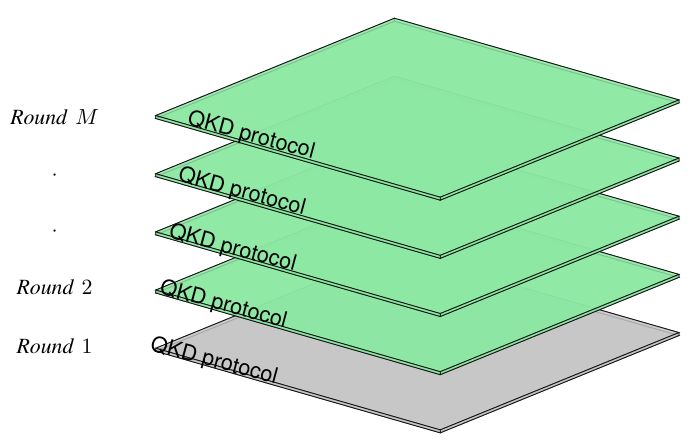}
		\caption{Block diagram of a round-wise QKD scheme.}\label{pc-qkd-scheme}
	\end{figure}

	%In the following, a description of a BB84-QKD process based on systematic polar coding is introduced in this section. A round-wise QKD process scheme is considered, see Figure \ref{pc-qkd-scheme}. Each round represents a single implementation (run) of a QKD protocol. \\
	\indent Polar-code QKD is a \textit{protocol process}\footnote{\textit{protocol process}\textemdash a series of $M$ protocol runs.} that consists of $M$ rounds ($M>>1$), see Figure \ref{pc-qkd-scheme} [\textit{Note}: Since polar coding can be performed not only in a binary alphabet (in the field $\mathbb{F}_2$) but also in a bigger alphabet (in a field $\mathbb{F}_a$, $a>2$) \cite{Arikan2009,Arikan2011}, Polar-code QKD can be realized by means of quantum systems of higher dimensions ($\dim(\mathcal{H})>2$, $\mathcal{H}$\textemdash Hilbert space of a quantum system)]. In each round (run), a QKD protocol is performed. In the following lines, description of a Polar-code QKD is presented.

	\begin{enumerate}
		\item[\textit{Round} $1$.] A BB84-QKD is executed in this round between Alice and Bob. The cryptographic key established in this round is used to "trigger" (play the role of a basis for) the next round. Therefore \textit{Round} $1$ can be interpreted as the \textit{initialization} of Polar-code QKD. Note that the key of this round is to be used for encrypting bits (frozen bits) of the next round.
		\item[\textit{Round} $2\divisionsymbol M$.] Each of these rounds is characterized by the following steps:
		\begin{enumerate}
			\item[\textit{Step} $1$.] Alice generates $K$ random key bits and prepares $F$ frozen bits (all the frozen bits are set to binary $0$). She performs systematic polar coding on the $K$ random bits and $F$ frozen bits according to Section \ref{polar-coding}. A sequence of $N$ bits is produced by the polar encoder ($N=K+F$). Note that both $K$ key bits and $F$ frozen bits enter the polar encoder, see Section \ref{polar-coding}. Alice encrypts the encoded $F$ frozen bits via the previous round key (key part).
			\item[\textit{Step} $2$.] Alice performs a random permutation on the $N$ bits. Alice prepares $N$ qubits in states corresponding to the $N$ encoded (and permuted) bits:
			\begin{equation*}
				0,1,0,\dots, 1 \rightarrow \ket{0}, \ket{1}, \ket{+}, \dots, \ket{-}
			\end{equation*}
			by using either the \textit{Hadamard} basis $\{\ket{+},\ket{-}\}$ or the \textit{Computational} basis $\{\ket{0},\ket{1}\}$, as can be seen in the above example. Alice picks the preparation basis (\textit{Hadamard} or \textit{Computational} basis) in a random and independent way for each qubit. Next Alice sends the $N$ qubits to Bob through lossless but noisy quantum communication channel.
			\item[\textit{Step} $3$.] Bob receives the $N$ qubits from Alice. Bob may use a \textit{non}-\textit{demolition} \textit{measurement} to verify the reception of each qubit. Alice announces the basis in which each qubit is prepared. Bob measures the received qubits in the corresponding bases (bases announced by Alice). Alice also announces the permutation she performed on the $N$ bits [\textit{Note}: The permutation of bits is equivalent to performing permutation on the qubits alone.]. Being aware of the permutation, Bob learns the positions of the $F$ frozen bits in the encoded sequence of $N$ bits [\textit{Note}: Recall that the encoded frozen bits are encrypted by Alice.]. Alice and Bob use the encoded frozen bits (or part of the encoded frozen bits)\footnote{This takes place if part of the encoded frozen bits is sufficient for parameter estimation to be performed properly.} to perform \textit{parameter estimation}. The parameter estimation consists of comparing the encoded frozen bits of Alice and Bob (the encoded bits prior to their transfer and encoded bits after their transfer) so that the \textit{quantum bit error rate} (QBER) of the quantum channel is evaluated. For this purpose, Bob announces his (received) encoded frozen bits (or part of his (received) encode frozen bits). Consequently Alice compares her encoded frozen bits to those announced by Bob. She thus finds (computes) the quantum bit error rate. Based on the value of this quantity, Alice decides whether or not this round is to be terminated\textemdash if the QBER exceeds a preliminary determined threshold, Alice and Bob terminate not only the current round but the whole QKD protocol process; otherwise Alice and Bob proceed forward to the next step. Note that Alice and Bob use for instance a threshold of $E = 4\%$ (or $E = 0.04$)\footnote{In order for Polar-code QKD to demonstrate satisfactory performance, the quantity $E$ needs to be less or equal to $0.04$ ($E \leq 0.04$), as shown in Section \ref{key-rate-perf}.} during the parameter estimation procedures performed in the framework of the proposed Polar-code QKD. This implies that the amount of frozen bits is $F = \lceil H(E)\cdot N \rceil = \lceil \approx0.24N \rceil$, see Section \ref{polar-coding}.
			\item[\textit{Step} $4$.] Bob decrypts (deciphers) his encoded frozen bits via previous round key (key part). Then he performs polar decoding on his $N$ bits. He obtains $K$ key bits out of the decoding ($\lceil H(E)\cdot N\rceil$ ($F$) amount of bits (the frozen bits) are discarded). Note that Bob's key bits are identical to those of Alice if they have properly performed each protocol procedure thus far. 
			\item[\textit{Step} $5$.] Alice and Bob perform privacy amplification procedure \cite{Bennett1995} (\textit{e.g.}, Toeplitz-based privacy amplification \cite{Yuan2018}). In this way, Alice and Bob establish a secure round key. Note that $\lceil H(e)\cdot N \rceil$ amount of bits are leaked (so subsequently discarded) in this procedure.  
			\item[\textit{Step} $6$.] Alice and Bob divide the established round key into two parts:
			\begin{enumerate}
				\item[(i)] part for encrypting frozen bits in the next round of the Polar-code QKD framework ($\lceil H(E)\cdot N \rceil$ amount of key bits);
				\item[(ii)] part for secure communication purposes ($N - \lceil H(e)\cdot N \rceil - \lceil H(E)\cdot N\rceil - \lceil H(E)\cdot N \rceil$ amount of key bits).
			\end{enumerate}
		\end{enumerate}
	\end{enumerate}
	Note that a part of \textit{Step} $3$ could be performed in the following way:\\
	\indent Alice reveals the positions of the encoded frozen bits (correspondingly, qubits). Using randomly and independently chosen measurement bases, Bob measures the qubits that correspond to the encoded frozen bits. Alice announces the initial encoded frozen bits and preparation bases of their corresponding qubits. Bob discards all his measurement results (his encoded frozen bits) for which his measurement bases mismatch the corresponding preparation bases of Alice. Bob compares his remaining measurement results (remaining encoded frozen bits) to the corresponding encoded frozen bits of Alice. In this way Bob evaluates the so-called quantum bit error rate. Bob uses the encoded frozen bits announced by Alice in the upcoming protocol procedures (steps).\\
	\indent \textit{Security of a Polar-code QKD}. Since the Polar-code QKD is identical to BB84 in terms of quantum stage, without loss of generality, one could assume that Polar-code QKD is as secure as BB84.\\
	
	\subsection{Key rate}\label{key-rate-perf}
\begin{figure}[h]
		\centering
		
		\includegraphics[scale=0.7]{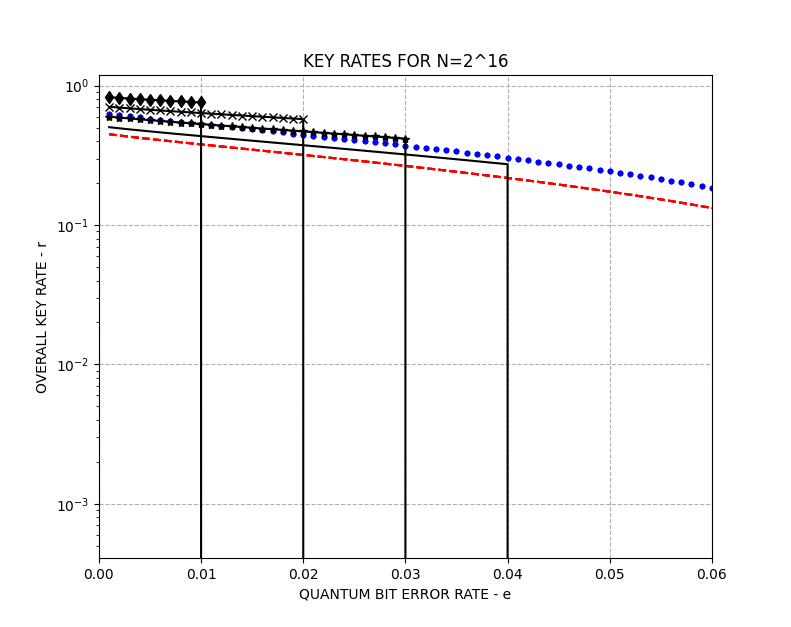}
		\caption{Key rate comparison between BB84-QKD ($r^{\text{BB84}}(e)$, dashed (red) line), Polar-code QKD ($r(e,E)$, solid (black) line), and eBB84-QKD ($r^{\text{eBB84}}(e)$, dotted (blue) line) in the case of $N=2^{16}$. The Polar-code QKD rate $r(e,E)$ is given for several values of quantity $E$: solid line\textemdash$E$ $=$ $0.04$; solid $*$ line\textemdash$E$ $=$ $0.03$; solid $\cross$ line\textemdash$E$ $=$ $0.02$; solid $\bLozenge$ line\textemdash$E$ $=$ $0.03$. Note that the BB84-QKD rate $r^{\text{BB84}}(e)$ does not depend on $E$. The infinite-slope ending of $r(e,E)$ is due to polar coding performance\textemdash the polar coding corrects errors up to error rates of $E$; \textit{i.e.}, Polar-code QKD does not operate for error rates above $E$.}\label{key-rate-comp}
	\end{figure}

\begin{figure}[h]
		\centering
		
		\includegraphics[scale=0.7]{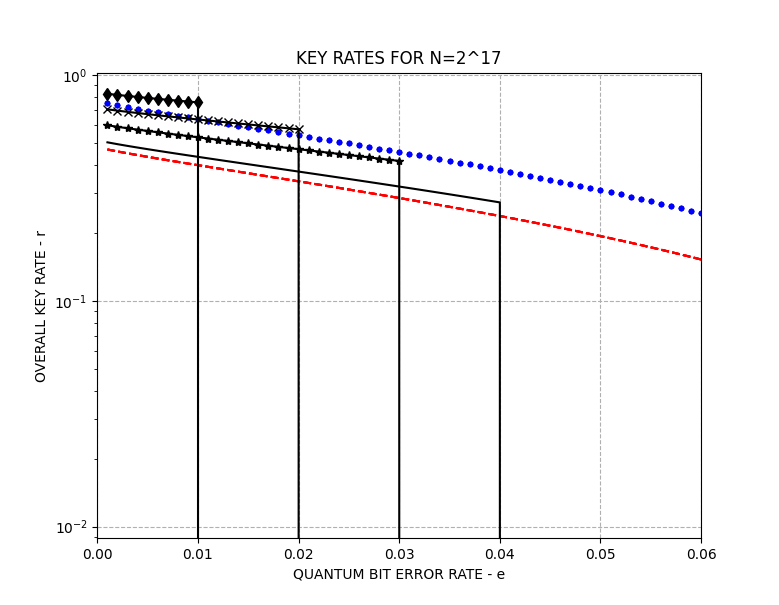}
		\caption{Key rate comparison between BB84-QKD ($r^{\text{BB84}}(e)$, dashed (red) line), Polar-code QKD ($r(e,E)$, solid (black) line), and eBB84-QKD ($r^{\text{eBB84}}(e)$, dotted (blue) line) in the case of $N=2^{17}$. The Polar-code QKD rate $r(e,E)$ is given for several values of quantity $E$: solid line\textemdash$E$ $=$ $0.04$; solid $*$ line\textemdash$E$ $=$ $0.03$; solid $\cross$ line\textemdash$E$ $=$ $0.02$; solid $\bLozenge$ line\textemdash$E$ $=$ $0.03$. Note that the BB84-QKD rate $r^{\text{BB84}}(e)$ does not depend on $E$. The infinite-slope ending of $r(e,E)$ is due to polar coding performance\textemdash the polar coding corrects errors up to error rates of $E$; \textit{i.e.}, Polar-code QKD does not operate for error rates above $E$.}\label{key-rate-comp2}
	\end{figure}

	To determine the key rate of the Polar-code QKD, the following expression is made used of \cite{Gottesman}
	\begin{equation}\label{round-key}
		R = s(1 - H(e) - fH(e))
	\end{equation}
	where $s$ is the \textit{sifting coefficient} ($s \in (0,1]$\textemdash$s=1$ corresponds to no sifting), $f$ is the efficiency of the error correction/information reconciliation procedure (usually $f=1.1\divisionsymbol1.2$), $e$ is the quantum bit error rate, and $H(\cdot)$ is the Shannon entropy. The above expression is used to evaluate the round key of a round-wise QKD scheme when an asymptotic regime of operation is considered ($N\rightarrow\infty)$. In the case of a finite-size regime (finite $N$), the above expression takes the form
	\begin{equation}\label{round-key-fs}
		R = s(1 - H(e) - fH(e) - \beta)
	\end{equation}
	where $0<\beta<1$ is a parameter that accounts for the amount of qubits being sacrificed for parameter estimation purposes. Note that $\beta$ depends on $N$ ($\beta \propto 1/N$).\\
	\indent The overall key rate of a round-wise QKD is given by
	\begin{equation}
		r = \frac{1}{M}\sum\limits_{i=1}^{M} R_i
	\end{equation}
	where $R_i$ is a round key rate (see Equation \eqref{round-key}) and $M$ is the number of rounds. In fact $r$ is the average round key rate of a round-wise QKD. The Polar-code QKD has a key rate of the form
	\begin{equation}
		r = \frac{1}{M}\left[-R_1 + \sum\limits_{i=2}^M R_i \right]
	\end{equation} 
	since the key established in the first round (trigger/initialization round) is sacrificed for second round purposes, namely encrypting encoded frozen bits. The round key rates $\{R_i; i\geq2\}$ are evaluated, as follows:\textemdash
	\begin{equation}\label{pc-round}
		R_i = R' = 1 - H(e) - \gamma - \zeta\; \; \; \;\;\; \text{given}\; i\geq2
	\end{equation}
	where $\gamma = H(E)$ corresponds to the amount of bits discarded during polar decoding (all the frozen bits are discarded) and $\zeta = H(E)$ corresponds to the amount of (round) key bits sacrificed for next round purposes. Note that $s=1$ in Equation \eqref{pc-round}\textemdash no sifting occurs in Polar-code QKD. In case of $M >> 1$ (or $M\rightarrow\infty$) the quantity $r$ becomes
	\begin{eqnarray}
		r &&= \lim\limits_{M\rightarrow\infty} \left[-\frac{R_1}{M} + \frac{1}{M}\sum\limits_{i=2}^M R_i\right] = \lim\limits_{M\rightarrow\infty} \left[-\frac{R_1}{M} + \frac{1}{M}(\approx M\cdot R')\right] \nonumber \\
		&& = \lim\limits_{M\rightarrow\infty} \left[-\frac{R_1}{M} + \frac{1}{M}(M\cdot R')\right] = \lim\limits_{M\rightarrow\infty} \left[-\frac{R_1}{M} + R'\right] = R'
	\end{eqnarray}
	where $R'$ is a round key rate for rounds $i\geq2$, as denoted above. Note that a round-wise BB84-QKD is characterized by a quantity $r$ of the form 
	\begin{equation}
		r^{\text{BB84}} = \frac{1}{M}\sum\limits_{i=1}^M R^{\text{BB84}}_i = \frac{M\cdot R^{\text{BB84}}}{M} = R^{\text{BB84}} = s\left(1-H(e)-fH(e)-\beta\right)
	\end{equation}
	where $s$ takes the value of $0.5$ \cite{Bennett1984}. The key rate of a round-wise efficient BB84-QKD (eBB84-QKD) is given by
	\begin{eqnarray}
		r^{\text{eBB84}} &&= \frac{1}{M}\sum\limits_{i=1}^M R^{\text{e BB84}}_i = \frac{M\cdot R^{\text{eBB84}}}{M} = R^{\text{eBB84}} = s\left(1-H(e)-fH(e)-\beta \right) \nonumber \\
		&&= 1-H(e)-fH(e) \;\;\;\;\;\; \text{given} \; N\rightarrow\infty
	\end{eqnarray}
	where $s$ tends to unity and $\beta$ to zero as $N$ tends to infinity ($s\rightarrow1$ given $N\rightarrow\infty$, asymptotic regime of a QKD protocol). For eBB84-QKD, $\beta$ is related to $p$, the probability of choosing \textit{Hadamard} basis for preparation or measurement purposes: $\beta=p^2$. Also the sifting coefficient $s$ is related to $p$ by the expression $s = 1 - 2p(1-p)$.\\
	\indent \textit{Key rate comparison}. In the key rate comparison, we consider that the QKD protocols operate in a finite-size regime. We examine the key rates of Polar-code QKD, round-wise BB84-QKD, and round-wise eBB84-QKD for $N = 2^{16}$ and $N = 2^{17}$. Also the key rate of the Polar-code QKD is evaluated for different values of $E$\textemdash error rate threshold in the parameter estimation procedure. We assume that $\sim5000$ bits (qubits) are sufficient for proper parameter estimation\textemdash an appropriate QBER resolution is supposed to be obtainable by comparing pairs of $\sim5000$ bits (qubits). This implies that $\beta N \approx 5000$ so that $\beta\approx0.08$ for $N = 2^{16}$ and $\beta\approx0.04$ for $N=2^{17}$. Then, for eBB84-QKD, we have $p=\sqrt{\beta}\approx0.283$ for $N=2^{16}$ and $p=0.02$ for $N=2^{17}$. Hence $s=(1-2p(1-p))=0.7$ for $N=2^{16}$ and $s=0.8$ for $N=2^{17}$. \\
	\indent In the following, we present a comparison between BB84-QKD, efficient BB84-QKD, and Polar-code QKD with regard to their key rates $r^{\text{BB84}}$, $r^{\text{eBB84}}$, and $r$, respectively. In fact the functions $r^{\text{BB84}}(e)$, $r^{\text{eBB84}}(e)$, and $r(e,E)$ are compared, as shown in Figure \ref{key-rate-comp} and Figure \ref{key-rate-comp2}. Note that Figure \ref{key-rate-comp} introduces a key rate comparison in the case of $N=2^{16}$ while Figure \ref{key-rate-comp2} introduces a key rate comparison in the case of $N=2^{17}$. Both figures display that Polar-QKD has a key rate advantage over BB84-QKD  and eBB84-QKD in the case of low error rates $e$. It is also evident that $r(e,E)$ increases as $e$ decreases\textemdash the key rate advantage of Polar-code QKD over BB84-QKD and eBB84-QKD becomes more prominent. Note that $r(e,E)$ is a finite function\textemdash it vanishes as $e$ reaches $E$, as depicted in Figures \ref{key-rate-comp} and \ref{key-rate-comp2}. This behavior is due to polar coding performance\textemdash the polar coding corrects errors up to error rates of $E$; \textit{i.e.}, Polar-code QKD does not operate for error rates above $E$.
	
	\section{Conclusion}\label{conclusion}
	The paper has introduced a quantum key distribution protocol process (a series of protocol runs) called Polar-code QKD. It is a scheme which combines two concepts, (i) quantum key distribution and (ii) systematic polar coding, in order for higher-rate quantum cryptography system to be obtained. The incorporation of polar coding into the quantum phase of a QKD provides a way to combine both parameter estimation and error correction procedures. This way the key-rate reduction due to classical error-correction communication is mitigated\textemdash this results in increasing the key rate. The key rate of the newly-proposed scheme has been defined and evaluated. As shown in the paper (Figures \ref{key-rate-comp} and \ref{key-rate-comp2}), Polar-code QKD achieves higher key rates than the original QKD (BB84-QKD) \cite{Bennett1984} and the so-called efficient standard QKD (eBB84-QKD) \cite{Lo2005}. Note that the advantage of Polar-code QKD is present when finite-size regime of operation and lower-error-rate quantum communication channel are taken into account.

	\section*{Acknowledgment}
	This work is supported by Technical University of Varna under Scientific Project Programme, Grant No. $\text{H}\Pi$6/2025. This research is also incorporated into the framework of Scientific Programme "Increasing National Scientific Capacity in the Field of Quantum Information", Bulgarian Ministry of Education and Science.

	\ifCLASSOPTIONcaptionsoff
	\newpage
	\fi

\end{document}